*instruments*

# π+π- decays of the $f_0(1370)$ scalar glueball candidate in pp Central Exclusive Production experiments

**Ugo Gastaldi [1]\* and Mirko Berretti [2]**

[1] INFN Sezione di Ferrara, Italy
[2] University of Helsinki and Helsinki Institute of Physics, Finland
\* Correspondence: gastaldi@cern.ch

**Abstract:** The study of the properties of π+π- pairs emitted in pp Central Production Experiments (CEP) shows that a) scalar and tensor mesons production dominates at increasing energies and that 2++ production is suppressed by selecting events with low four momentum transfer square |t| at both proton vertices and b) π+π- decays of the $f_0(1370)$ meson are directly observable as an isolated peak over a vanishing background between 1.1 and 1.6 GeV in measurements at high energies and low |t|. Together with the measurements of the decay branching ratios of the $f_0(1370)$ into σσ, ϱϱ, ππ, KKbar, ηη from pbar annihilation at rest, these observations point at a high glue content of the $f_0(1370)$ and suggest that LHC and central detectors supplemented by precision forward detectors installed in roman pots could be used as a unique clean source of all the low energy scalars.

**1. Introduction**

The possibility of the existence of glueballs is a basic qualitative prediction of quantum chromo dynamics (QCD)[1,2]. The lowest lying glueball states are expected to have the same 0++ JPC quantum numbers as the vacuum. While for the 0-+ pseudoscalar, 1-- vector and 2++ tensor meson nonets there are two observed isoscalar partners for two places in each nonet (respectively η and η', ω and φ, $f_2(1270)$ and $f'_2(1525)$, there are 4 or 5 observed isoscalar mesons (respectively σ, $f_0(980)$, $f_0(1370)$, $f_0(1500)$, $f_0(1710)$ for the two places of the isoscalar members of the 0++ ground state nonet. The candidates for the two places are 4 or 5 depending whether σ and $f_0(1370)$ are considered as distinct separated objects or they are part of a single continuum. The existence of the σ meson is considered established since some time (see [3] for a review) and σ is currently called $f_0(500)$. Doubts instead concern the existence of the $f_0(1370)$ as an individual isolated structure (see [4] for a review). Concerning the nature of the 0++ mesons, various scenarios envisaged in the literature include (because of mixing) the possibility of a qqbar, qqqbarqbar, qqbar-qqbar and gg content in their wave functions. If the 0++ isoscalars are 5 it is more difficult to exclude the hypothesis of the presence (or even dominance) of a gg content in some of them. Production in glue rich processes (central production mediated by double pomeron exchange, pbar annihilations, J/ψ and ψ' radiative decays, heavy meson decays) and absence in γγ production and meson exchange are criteria useful to characterize the gg content of the scalars [5]. Relative decay branching ratios into σσ, ϱϱ, ππ, KKbar, ηη for the three heavier scalars, and also into ηη' for $f_0(1500)$ and $f_0(1710)$, give other selection criteria to identify their gg content [5-8]. Currently there is no consensus concerning the experimental observation of a scalar glueball nor on the possible gg content in the scalars experimentally observed (for reviews see [4,9-14]).

This talk is focused on the production of π+π- pairs in pp Central Exclusive Production (CEP). CEP experimental data have been scrutinized in function of several dependences (s^1/2 pp cms energy, four momentum transfer square |t| at both proton vertices, angular correlations of the two scattered proton transverse momenta) in ref [15] (which gives an extensive reference list). From the ensemble of the pp CEP data it has emerged that higher energy favors 0++ and 2++ meson production, as expected because of the dominance of pomeron-pomeron exchange [5,16], and that restricting to data at low |t| depresses 2++ production, so enhancing evidence of 0++ production. By combining the results of the analysis of the data of the AFS experiment [17,18] at ISR, the energy dependence of double-pomeron, pomeron-reggeon and reggeon-reggeon production in CEP [5,16] and data of the STAR experiment [19-22] at RICH, it has emerged [15] that the $f_0(1370)$ scalar meson is the main responsible for an isolated peak -positioned between 1.1 and 1.6 GeV and well separated from the $f_0(980)$ signal- which is observable in low |t| STAR data [19,20]. We focus here the discussion on the STAR data and their interpretation. Evidences for the $f_0(1370)$ as a confined structure observable in other two pseudoscalar decay channels and into 4 pion decays are also mentioned and are covered more extensively in refs [23,24]. Prospects for the spectroscopy of low mass $0^{++}$ scalars at LHC and consequences of the assessment of the existence of $f_0(1370)$ for the observation of a scalar glueball are discussed in the final sections of this paper.

## 2. $f_0(1370)$ $\pi^+\pi^-$ decays in pp CEP data

Fig. 1 [19] reproduces a $\pi^+\pi^-$ mass plot of pp CEP data of the STAR experiment at RICH collected at $\sqrt{s}$ = 200 GeV in 2009. In the 2009 run the machine optics was optimized with ß*≈ 20 m for a low |t| kinematical coverage necessary for measurements of the pp elastic cross section. Two sets of detectors of the two forward scattered protons were installed in roman pots on each side of the interaction region at 55.5 and 58.5 meters distance from the interaction region. The central detector which surrounds the interaction region was active and preliminary $\pi^+\pi^-$ CEP spectra were presented in 2012 and 2014 [19,20].

The spectrum of fig.1 contains 588 events selected in the kinematical range : 0.003< |t$_1$|,|t$_2$| < 0.03 GeV$^2$ ( |t$_1$| and |t$_2$| are the four momentum transfer squared at the two proton vertices (it is worth noticing that the four momentum transfer acceptance window of these CEP data features the lowest |t$_{min}$| of all existing CEP experiments); pseudorapidity of single pions |η$_\pi$|<1.0, transverse momentum balance of the two pions and two protons present in the final state p$^t_{miss}$ < 0.02 GeV. The spectrum of fig.1 is not corrected for acceptances nor detection efficiencies. However in the selected kinematical range the acceptance of all the variables exceeds 10%. The background represented by the spectrum of like sign pion pairs contains very few events and it is featureless. Notice then that the absence of events in the 1-1.2 GeV energy window is not due to an effect of acceptances. Fig.2 [20] shows the spectrum corrected for acceptances.

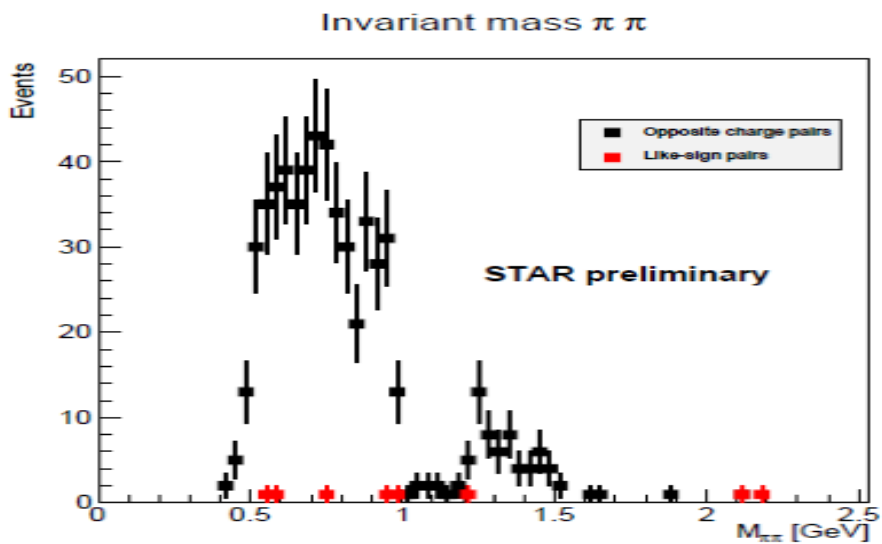

**Figure 1.** $\pi^+\pi^-$ mass plot of STAR pp CEP data of the 2009 run with 0.003<|t$_1$|, |t$_2$|<0.03 GeV$^2$ kinematic coverage of both scattered protons: row data [ 19 ]. Fig. from ref.[19].

For the 2015 STAR run the two sets of roman pots containing the proton detectors on the two sides of the central detector were approached to the interaction region at distances of 15.8 and 17.6 meters and the machine optics was set with ß*≈ 0.85 m in order to increase the |t| acceptance [21]. The |t| window at the two proton vertices was then : 0.03< |t$_1$|,|t$_2$| < 0.3 GeV$^2$. The $\pi^+\pi^-$ mass spectrum of the 2015 STAR pp CEP run is reproduced in fig. 3 [21,22].

The STAR spectrum of fig.3 collected with high |t| windows features a signal raising from threshold to 0.9 GeV and a sharp drop across the $f_0(980)$ by a factor of about 5 between the maximum at 0.9 GeV and the minimum at 1.1 GeV. A strong $f_2(1270)$ signal dominates the 1.1-1.5 GeV region. The ordinate of the signal in the valley which separates at 1.1 GeV the $f_0(980)$ drop

from the $f_2(1270)$ peak is about 1/4 of the $f_0(980)$ peak value. The activity above 1.6 GeV is at a level of about 1/5 of the value of the minimum at the 1.1 GeV valley.

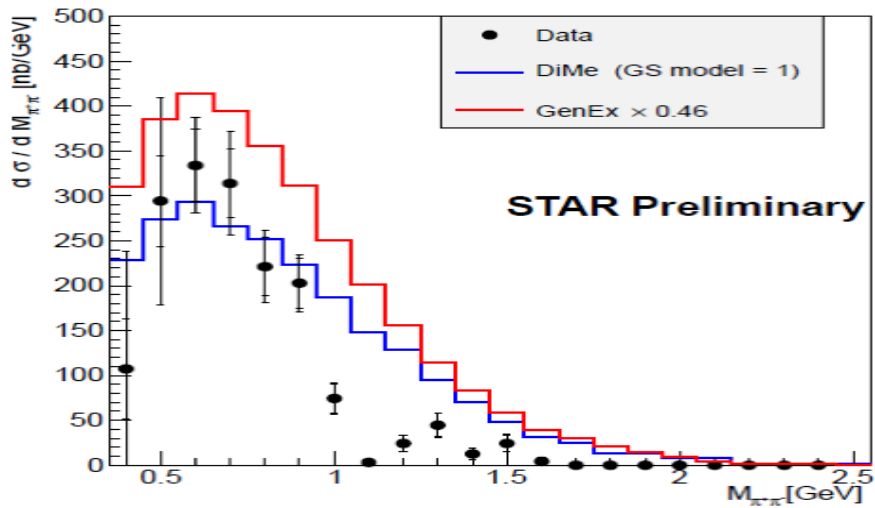

**Figure 2.** π+π- mass plot of pp STAR data of the 2009 run with 0.003˂|t₁|, |t₂|˂0.03 GeV² kinematic coverage of both scattered protons: data corrected for acceptance [20]. Fig. from ref. [20].

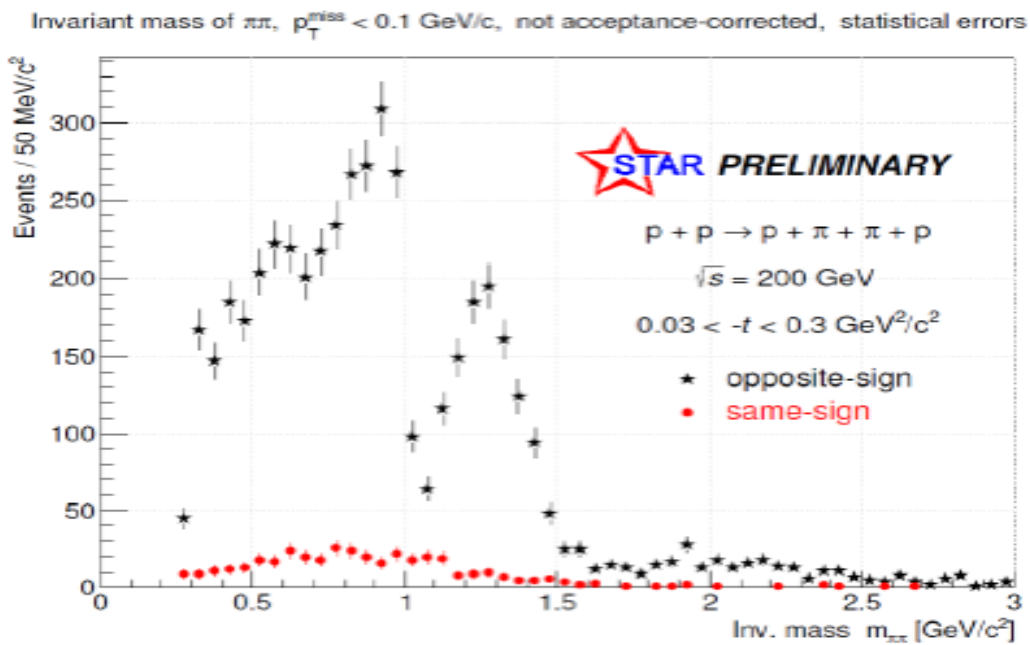

**Figure 3.** π+π- mass plot of STAR ppCEP data of the 2015 run with 0.03˂|t₁|, |t₂|˂ 0.3 GeV² kinematic coverage of both scattered protons [21]. Fig. from ref.[21].

The STAR spectrum of fig.1, collected with the lowest |t| windows ever explored in CEP experiments, features a strong σ signal reminiscent of the σ signal observed in the AFS experiment at the CERN ISR (see fig. 4 from ref. [17]. The σ signal is followed by a sharp drop between 0.9 and 1.1 GeV. A completely isolated structure is present in the window 1.2-1.5 GeV , it is centered at about 1.35 GeV and is generated by about 60 events. The ratio between the values of the ordinates at 0.9 and at 1.1 GeV is above 15 and exceeds the value of about 10 in the AFS data [17,18]. As already mentioned above, the valley around 1.1 GeV is not an acceptance nor a detection efficiency effect, because the acceptance and detection efficiencies vary smoothly across the full energy window relevant for σ, $f_0(980)$, $f_0(1370)$, $f_0(1500)$ and $f_0(1710)$. The STAR structure in the energy window 1.1-1.6 GeV has been interpreted [15] as due to dominant production of $f_0(1370)$ for the reasons listed in the following.

1) The analysis of the angular distributions of the AFS data at 63 GeV total energy has shown that S-wave production dominates up to 1.7 GeV with a small contribution of D-wave, associated to f$_2$(1270) production [17,18,13].
2) The S-wave contribution associated to double-pomeron exchange is expected to remain constant with increasing $\sqrt{s}$ , while the contributions associated to pomeron-reggeon and reggeon-reggeon exchange are expected to decrease as $s^{-1/2}$ and $s^{-1}$ [5,16].
3) The $2^{++}$ contribution (also due to double-pomeron exchange) is also expected to remain constant with increasing $\sqrt{s}$, however the f$_2$(1270) signal reduces dramatically at fixed energy when the events are selected by restricting the four momentum transfer ltl at both proton vertices to the lowest values, as witnessed by the STAR data themselves by comparing fig.1, where f$_2$(1270) is not visible, with fig.3, were it is a dominant feature, and as witnessed at experiments at ISR at 63 GeV, were CEP data were collected with different values of the ltl windows (exemples and references are given in ref.[15]. The ltl acceptance of the AFS data was 0.01˂ltl˂0.06 at both proton vertices. The ltl acceptance of the STAR events of fig.1 is partly inside and partly extends below the low ltl AFS region.
4) Because of points 1 and 2 the S-wave contribution in fig.1 is dominant in the energy region 1.1-1.6 GeV and, because of point 3, the D-wave contribution in the STAR data (if any) is reduced in comparison to that in the AFS data of fig.4.
5) The structure in fig.1 between 1.1 and 1.6 GeV and centered around 1.35 GeV is too large to be associated to the f$_0$(1500), which has a width of the order of 100 MeV, and is expected at 1.5 GeV. The statistics is too low to extract meaningfully a ratio between the f$_0$(1370) and the f$_0$(1500) contributions. However most of the events occur at masses below 1.5 GeV, so that, as indicated in the analysis of the AFS data [17,18] the f$_0$(1500) may contribute to the spectrum via a destructive interference with the f$_0$(1370) amplitude. A little contribution of f$_2$(1270) might also be present, but it looks marginal because the low energy tail of the f$_2$(1270), which extends in the region below 1.2 GeV in fig. 3, is not present in the spectrum of fig. 1, where the region below 1.2 GeV is nearly empty.

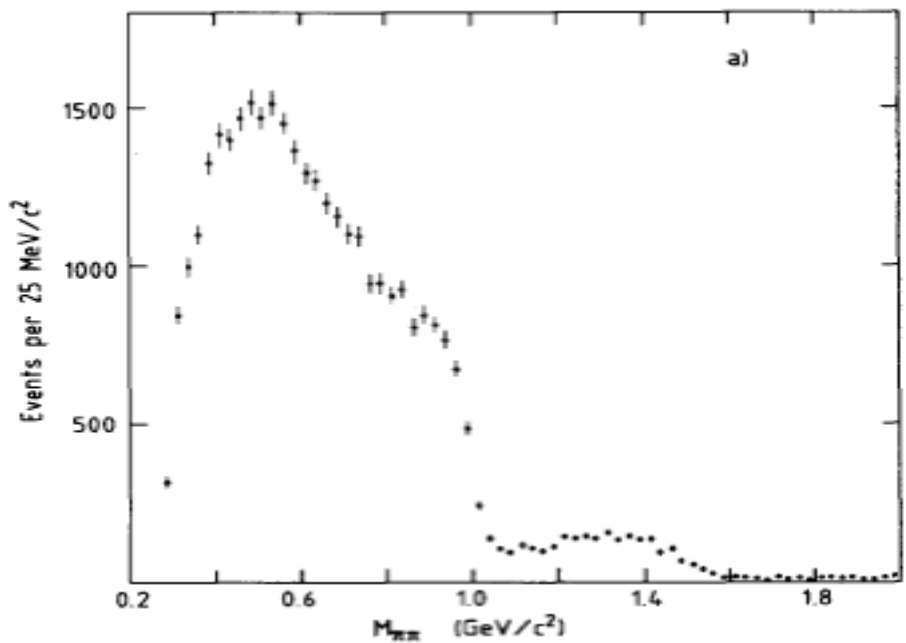

**Figure 4.** π$^+$π$^-$ mass plot from pp CEP interactions at $\sqrt{s}$ =63 GeV measured by the AFS experiment at ISR : raw data [17]. Fig. from ref. [17].

Independently of the interpretation of the structure in the 1.1-1.6 GeV energy region, quite noticeable is the fact that the STAR π$^+$π$^-$ spectrum of fig.1 drops nearly to zero before 1.1 GeV. This may be the result of the interference of the amplitudes of the low energy tail of the f$_0$(1370) with the high energy part of the f$_0$(980) plus the effect of the KKbar threshold, but very likely it might be due to the vanishing of the S.wave continuum for events selected in the low ltl kinematical region. The S-wave continuum, which is usually invoked with its destructive interference with the $f_0(980)$ amplitude to generate the drop at 1 GeV, seems drastically reduced in comparison to the AFS data, which were collected with ltl at both proton vertices comprised between 0.01 and 0.06 GeV$^2$. It looks like the σ meson, which generates the broad structure above 0.5 GeV, is confined below 1 GeV. In other words, under this hypothesis the "red dragon" glueball proposed by Minkowski and Ochs [25], which features a low energy body centered at about 0.6 GeV and a head extending below the $f_0(1500)$, would be split into two separate parts, the σ and the relatively narrow $f_0(1370)$.

## 3. Evidences for $f_0(1370)$ in other 2 pseudoscalar decay channels and in 4 pion decays

Evidence for the existence of the $f_0(1370)$ scalar as a structure of limited width (200-300 MeV) and mass centered near 1370 MeV is present since long in data of pbar annihilations which are not quoted in the PDG compilations or which are quoted and used in the $f_0(1500)$ section of the PDG compilations of the last two decades to produce average values for mass and width of $f_0(1500)$, but are discarded in the $f_0(1370)$ section.

Decays of $f_0(1370)$ into $K^+K^-$ and into $K_sK_s$ are directly observable in $\pi^0K^+K^-$, $\pi^-K_sK_s$, $\pi^0K_sK_s$ Dalitz plots and in the respective $K^+K^-$ and $K_sK_s$ mass plots of pbar annihilations at rest in liquid $H_2$ and $D_2$ targets (see fig. 5 from ref. [26], data from refs. [26-34]).

$K^+K^-$ decays of $f_0(1370)$ are the main source of the diagonal $K^+K^-$ band comprised between the $K^{*+}$ and $K^{*-}$ bands in the $\pi^0K^+K^-$ Dalitz plot of the top left plate of fig.5 and of the peak at 1.4 GeV in the $K^+K^-$ mass plot of the top right plate of fig. 5.

Destructive interference of the $f_0(1500)$ amplitude with the $f_0(1370)$ amplitude generates the deep narrow flat valley at $M(K_sK_s)=1.5$ GeV in the $\pi^0K_sK_s$ Dalitz plot and in the $K_sK_s$ mass plot obtained [26] using CERN [29,30] and BNL [31] bubble chamber data (see the bottom left and right plates in fig. 5). The same effect is visible in the high statistics $\pi^0K_lK_l$ Dalitz plot and in the $K_lK_l$ mass plot produced by Crystal Barrel [32] ( see figs.1 and 2 in [32]), where the valley is half as deep because of background and less mass resolution.

In the $\pi^-K_sK_s$ Dalitz plot produced [26] using CERN [33] and BNL [34] bubble chamber data (see the central left plate in fig. 5) is clearly noticeable the effect of the interference of the $f_0(1370)$ band with the $K^{*-}$ bands. The $K_sK_s$ decays of $f_0(1370)$ are the main source of the peak at about 1.4 GeV in the $K_sK_s$ mass plot of the central right plate of fig.5.

Decays of $f_0(1370)$ into $\pi^+\pi^-\pi^+\pi^-$ and into $4\pi^0$ measured in $\pi^-\pi^+\pi^-\pi^+\pi^-$ [35,36] and into $\pi^-4\pi^0$ [37,38] pbar annihilations at rest in liquid $D_2$ are the dominant feature (well distinct from phase space) in the respective mass plots (see fig. 6 left plate from ref. [36] and fig. 6 right plate from ref.[37]). These data show that $f_0(1370)$ decays dominantly into σσ (with a frequency twice that of the ϱϱ decays and 6 times larger than ππ decays) and that 4 pion decays of $f_0(1370)$ are 10 times more frequent than 4 pion decays of $f_0(1500)$ [37].

The fact -clearly confirmed by the STAR $\pi^+\pi^-$ decays of $f_0(1370)$- that $f_0(1370)$ has limited width and its mass is centered around 1370 MeV validates the analysis of data of pbar annihilations at rest into 3 pseudoscalars that used the hypothesis of the existence of the $f_0(1370)$ as an individual object [28 ,39,40]. The $f_0(1370)$ decay branching ratio into $\pi^+\pi^-$ of previous analysis is consequently reduced and the ratios of the decay branching ratios of $f_0(1370)$ into σσ, ϱϱ, ππ, KKbar and ηη to the decay branching ratios of $f_0(1370)$ into ππ are then of the order of 5.6, 2.9, 1, 1 and 0.02 [10,24].

The scenario which emerges is compatible with the expectation for a glueball to decay with equal rates into ππ and KKbar deriving from the equal couplings of gluons to nnbar and ssbar quarks [1,5] and expectations from QCD sum rules and low energy theorems for a scalar glueball [41]. There is however the caveat that in the analysis of WA102 pp CEP data at 29 GeV cms the ϱϱ decay branching ratios of f$_0$(1370) are more than 4 times larger than the σσ decay branching ratios [42] and the ratios of the decay branching ratios of f$_0$(1370) into ππ, KKbar and ηη to the decay branching ratios of f$_0$(1370) into ππ are of the order of 1, 0,46 and 0,16 [43,44]. These discrepancy should be clarified with CEP data at LHC. The assessment of the values of the branching ratios of decays of the f$_0$(1370) scalar meson into σσ and ϱϱ pairs and into ππ, KKbar, ηη pairs of pseudoscalar mesons derived from pbar annihilations at rest is discussed in more detail in ref. [24].

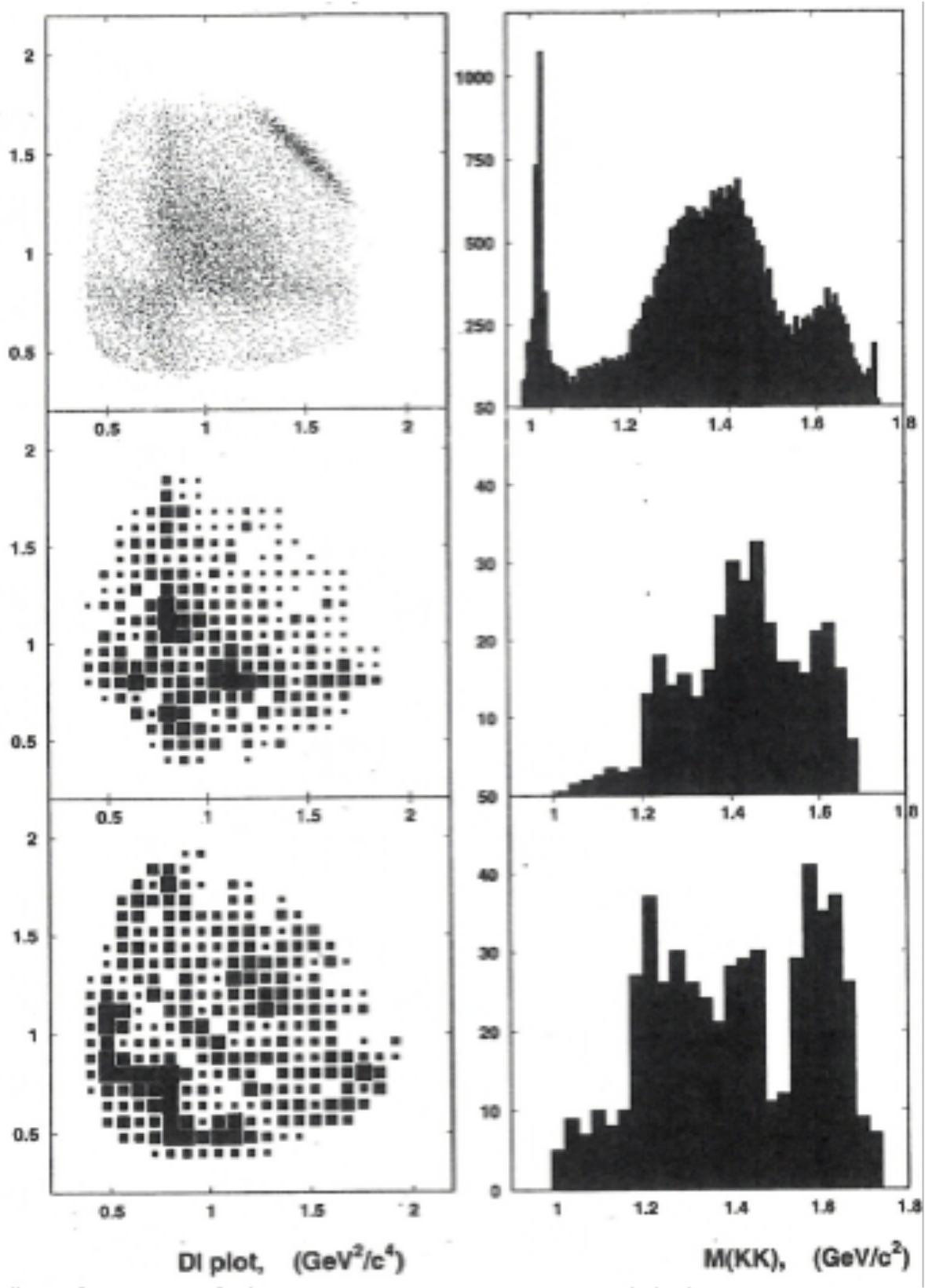

**Figure 5.** π⁰K⁺K⁻ , π⁻KₛKₛ , π⁰KₛKₛ Dalitz plots (top, middle and bottom left plates) and respective K⁺K⁻ and KₛKₛ mass plots (right plates) of pbar annihilations at rest in liquid H₂ and D₂ targets (picture derived from figs. 1, 2 and 5 of ref. [26].

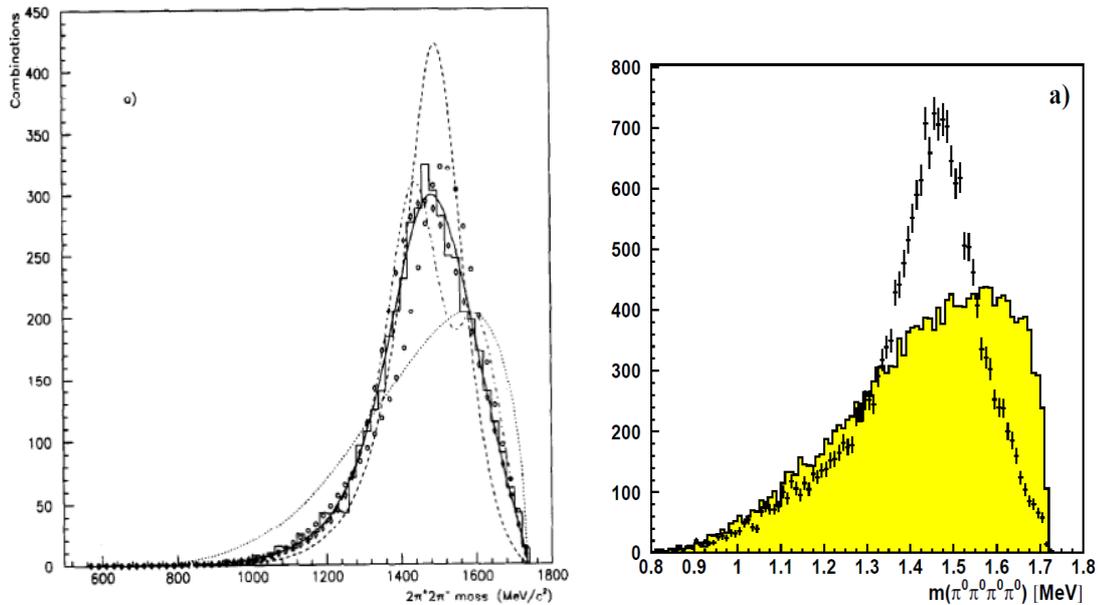

**Figure 6.** π+π- π+π- invariant mass in npbar → π-π+π- π+π- annihilations in liquid D$_2$ ;the histogram shows data, the dotted line shows phase space (left plate, from ref.[36]). 4π$^0$ invariant mass in npbar → π- 4π$^0$ annihilations at rest in liquid D$_2$; the peak –data points with errors- is mainly due to f$_0$(1370) decays to 4π$^0$, the colored histogram shows phase space (right plate, from ref. [37]).

## 4. Prospects for low mass scalars at LHC

At the higher LHC energies, non pomeron-pomeron exchange will be further depressed in CEP reactions. This will result in mass spectra with dominant 0$^{++}$ and 2$^{++}$ contributions. Extrapolating from π+π- AFS and STAR data, pp CEP data at LHC with low ltl at both proton vertices should possibly feature π+π- mass spectra with separated peaks of all the low energy scalars.

In particular the f$_0$(1370) peak should be free of f$_2$(1270) contamination by organizing runs that would enable to take CEP data with ltl at both proton vertices below 0.01GeV$^2$. If the suppression of 2$^{++}$ contributions at low ltl will be confirmed also for KKbar spectra, it should be possible to observe and measure f$_0$(1370) decays into K+K- and K$_s$K$_s$ without contamination of f$_2$(1270) decays and to compare directly π+π- , K+K- and K$_s$K$_s$ f$_0$(1370) decays. Likewise it should be possible to observe and measure K+K- and K$_s$K$_s$ f$_0$(1500) decays without contamination of f'$_2$(1525) events.

The relative decay branching ratios into π+π- , K+K- and K$_s$K$_s$ of the 3 scalars of higher mass, f$_0$(1370), f$_0$(1500), f$_0$(1710) would then be measurable in a reliable way. Also the study of the decays of f$_0$(1370), f$_0$(1500) and f$_0$(1710) into 4 charged pions would be simplified by the absence of contamination of decays of non 0$^{++}$ mesons. A critical point is to establish the ratios between their σσ and ϱϱ decays. Another critical point is to understand the narrow π+π- π+π- peak observed by WA76[45], WA91[46,47] and WA102[48] experiments near 1.45 GeV and tentatively attributed to interference of the f$_0$(1370) and f$_0$(1500) amplitudes.

In order to measure events with very low ltl it is necessary that the detectors of the two scattered protons be very near to the circulating beam. This makes the use of roman pots indispensable.

In LHC runs with β*=90 m, events with ltl as low as 0.03 GeV$^2$ can be measured with the apparata of the ATLAS-ALFA and CMS-TOTEM collaborations, which are equipped with precision detectors installed in roman pots 220 m upstream and downstream of the central detectors[49]. This kinematical region intercepts the ltl window of the AFS data and is just above the ltl acceptance window of the 2009 STAR data. The 2015

LHC β*= 90m data should therefore permit at least to confirm the results extracted from the π+π- STAR data and extend measurements to KKbar spectra.

## 5. Conclusions

The low |t| 200 GeV 2009 STAR π+π- spectrum features an isolated peak between 1.1 and 1.6 GeV disconnected from the signal drop at 1 GeV associated to the $f_0(980)$. The main source of the peak is the $f_0(1370)$ scalar. This conclusion is based on the S-wave dominance up to 1.7 GeV measured in the π+π- spectrum by the AFS experiment at 63 GeV, on the scaling with energy of the interactions which contribute to CEP data and on the observed disappearance of D-wave when the |t| window is restricted to low |t| values. Furthermore the π+π- continuum is interrupted around 1.1 GeV. It is likely that the continuum be dependent on |t|. If the low |t| STAR π+π- spectrum is interpreted as resulting from the contribution of π+π- structures and of a continuum, the continuum stops before 1.1 GeV and the peak between 1.1 and 1.6 GeV is due mainly to the $f_0(1370)$ with a minor contribution of $f_0(1500)$. The amplitude of the $f_0(1500)$ would interfere destructively with that of $f_0(1370)$. It is confirmed that the $f_0(1370)$ is a well identified structure with width of the order 200-300 MeV and centered around 1370 MeV. K+K- and $K_sK_s$ decays of $f_0(1370)$ also generate structures not larger than 200-300 MeV well visible in Dalitz plots of pbar annihilations at rest. The ensemble of these informations supports and validates the results of analyses of LEAR data of pbar annihilations at rest which lead to the conclusion that the ratio of ππ to KKbar decays of $f_0(1370)$ is of the order of 1.

The $f_0(1370)$ structure gives the largest contribution in pbar annihilations at rest into 4 pions. The π+π-π+π- $f_0(1370)$ signal has first been observed in CERN bubble chamber data of pbar annihilations in liquid $D_2$ in 1966 [35]. Reanalysis in 1993 of the π+π- π+π- $f_0(1370)$ mass plot and angular distributions of BNL bubble chamber data of pbar annihilations in liquid $D_2$ resulted in establishing the $0^{++}$ $J^{PC}$ quantum numbers of the $f_0(1370)$ and that σσ is its dominant decay channel [36]. Crystal Barrel at LEAR has measured npbar → π- $4π^0$ and npbar → π- $2π^0π^+π^-$ at rest in liquid $D_2$ and ppbar → $5π^0$ and ppbar → $3π^0π^+π^-$ annihilations at rest in liquid $H_2$. The analysis of these data has evolved and improved over several years [ 50,51,37,38,10 ]. Other resonances were found to contribute to the 4 pion spectra. A stable result through the evolution of the analysis has been that $f_0(1370)$ gives the dominant contribution, its main decay (contributing to the 4 pion spectra) is σσ; σσ decays are more frequent than ϱϱ decays by a factor 2. Furthermore also for $f_0(1500)$ σσ decays are more frequent than ϱϱ decays by a factor 2[37,10].

The production and decay properties of $f_0(1370)$, which emerge from the above experimental situation, match rather well general glueball characteristics [5] and expectations based on QCD sum rules and low energy theorems for a scalar glueball [41]. However the results of the analysis of CEP experiments at 29 GeV by the WA102 experiment [42-44] concerning 4 pion decays of $f_0(1370)$ and $f_0(1500)$ is in strong disagreement with the output of pbar annihilation experiments at rest: for WA102 $f_0(1370)$ σσ decays are less frequent than ϱϱ decays by a factor exceeding 4; $f_0(1500)$ σσ decays are at most 80% of ϱϱ decays [42].

LHC has the potential to produce in pp CEP reactions at very low |t| mass spectra of two pseudoscalars with separated $0^{++}$ peaks free from $2^{++}$ contamination. This should in particular permit to resolve the conflicts concerning the $f_0(1370)$ decay branching ratios into σσ and ϱϱ and clarify the scenario of all the scalar glueball candidates.

Precision detectors positioned inside roman pots are indispensable to measure events at very low |t|. LHC running with lattice configurations giving access to very low |t| and with central detectors tuned to measure events with low energy charged and also neutral particles would represent a factory for low energy scalars. This could establish the existence of gluonium and measure the distribution of gg in the wave function of all low energy scalars.